\documentstyle[emulateapj,psfig]{article}
\input{psfig}

\begin{document}

\title{Constraints on the Collisional Nature of the Dark Matter\\ from
Gravitational Lensing in the Cluster A2218}

\author{Priyamvada Natarajan$^{1,2}$, Abraham Loeb$^{3}$, Jean-Paul 
Kneib$^{4}$ \& Ian Smail$^{5}$}

\affil{1 Department of Astronomy, Yale University, New Haven, CT, USA}
\affil{2 Institute of Astronomy, Madingley Road, Cambridge, CB3 0HA, UK}
\affil{3 Department of Astronomy, Harvard University, 60 Garden Street, 
Cambridge, MA 02138, USA}
\affil{4 Observatoire Midi-Pyrenees, 14 Av. E.Belin, 31400 Toulouse,
France}
\affil{5 University of Durham, Department of Physics, South Road, 
Durham DH1 3LE, UK}

\begin{abstract}
The detailed distribution of mass within clusters of galaxies can
be used to probe the nature of dark matter. We show that constraints 
on the extent of the mass distribution around galaxies 
in the rich cluster A\,2218 
obtained from combining strong and weak lensing observations are 
consistent with the predictions which assume that the dominant
mass component (dark matter) in these halos is collisionless. 
A strongly interacting (fluid-like) dark matter is ruled-out at a 
confidence level of more than $5\sigma$.  
\end{abstract}


\section{Introduction}

The detection of gravitational lensing of distant galaxies by
foreground galaxies has been used in recent years to set constraints
on the masses and sizes of the lensing galaxy halos. Current studies
suggest the presence of halos extending beyond 100\,kpc (Brainerd,
Blandford \& Smail 1996; Griffiths et al.\ 1996; Dell'Antonio \& Tyson
1996; Fischer et al.\ 1999; McKay et al.\ 2001;
Wilson et al.\ 2001). The same technique can also be applied to
galaxies in the dense environments of massive X-ray clusters. Analysis
of galaxy-galaxy lensing in the cores of rich clusters implies a
smaller spatial extent for the dark matter halos associated with
morphologically-classified, early-type cluster members than that seen
in the field (Natarajan et al.\ 1998).  Most recently, Natarajan,
Kneib, \& Smail (2002) [NKS02 hereafter] have found compelling
evidence for the tidal truncation of dark matter sub-halos in clusters
based on {\it Hubble Space Telescope} imaging of lensing clusters at
intermediate redshifts.

The detailed mass distribution within lensing X-ray clusters, and in
particular the fraction of the total cluster mass that is associated with
individual galaxies -- has important implications for the frequency and
nature of galaxy interactions. As we show in this {\it Letter}, the
characteristic halo sizes of cluster galaxies that survive tidal
deformation and stripping in their dense environments offer tantalizing
clues as to the nature of dark matter itself. While there is
incontrovertible evidence for the existence of significant amounts of dark
matter in the Universe, its nature remains one of the challenging and
unsolved problems in cosmology. Following the original suggestion by
Furlanetto \& Loeb (2002), we demonstrate here that constraints on the
properties of the dark matter, whether it is collisionless or collisional
(fluid-like), can be derived based on the truncation radii of galaxy halos
in clusters.

If the dark matter is collisionless, then the sizes of galaxy halos in
clusters are primarily shaped by the dynamical effects of tidal truncation
and collisional stripping (Ghigna et al.\ 1998).  The dominant process of
tidal truncation acts on the short orbital time-scale due to the tidal
field of the cluster as a whole (Taylor \& Babul 2001), while collisional
stripping results from binary interactions among individual galaxies
(Binney \& Tremaine 1987).  The global tidal field of the cluster truncates
the dark matter halo of each galaxy at a radius inside of which the mean
mass density of the galaxy is roughly equal to the mean interior density of
the cluster. However, if the dark matter is fluid-like then galaxy halos
would be further stripped by ram pressure down to a radius which is
significantly smaller than the tidal radius (roughly by the ratio between
the velocity dispersion inside the galaxy halo and the velocity of the
galaxy through the cluster; see Furlanetto \& Loeb 2002 for further
details). In this work, we use the truncation radii of a sample of cluster
galaxies, inferred observationally from the galaxy-galaxy lensing in the
cluster A\,2218 (NKS02), to decide whether the dark matter is collisionless
or fluid-like.

\section{Using Galaxy-Galaxy Lensing to Measure Truncation Radii of 
Cluster Galaxies}

We model the background cluster as a superposition of a smooth large-scale
potential ($\sim 250^{\prime\prime}$ in extent) and small-scale clumps that
are associated with bright early-type cluster galaxies (see Natarajan \&
Kneib 1997 and Natarajan et al.\ 1998, for more details).  To quantify the
lensing distortion induced by the global potential, both the smooth
component and individual galaxy-scale halos are modeled self-similarly
using a surface density profile, $\Sigma(R)$, which is a linear
superposition of two pseudo-isothermal elliptical components (see the PIEMD
models derived by Kassiola \& Kovner 1993),
\begin{eqnarray}
\Sigma(R)\,=\,{\Sigma_0 r_0 \over {1 - r_0/r_t}} \left({1 \over
\sqrt{r_0^2+R^2}}\,-\,{1 \over \sqrt{r_t^2+R^2}}\right),
\end{eqnarray}
with a core radius $r_0$ and a truncation radius $r_t\,\gg\,
r_0$. The free parameters of this profile are chosen for both the smooth
component and the clumps so as to obtain the appropriate mass
distributions on the relevant scales. The projected radius $R$ is a
function of the sky coordinates $x$ and $y$ and the ellipticity $\epsilon$
(see \S 2.2 of Natarajan \& Kneib 1997). One of the attractive features of
this model is that the total mass is 
finite
($\propto \,{\Sigma_0} {r_0} {r_t}$). With the additional assumption that
light traces mass, galaxy halos in clusters are characterized by the
following scaling laws:
\begin{eqnarray}
{\sigma}\,=\,{\sigma_{*}}\left({L \over L_*}\right)^{1 \over 4};\,\,
{r_0}\,=\,{r_{0*}}{\left({L \over L_*}\right) ^{1 \over 2}};\,\,
{r_t}\,=\,{r_{t*}}{\left({L \over L_*}\right)^{\alpha}}.
\label{eq:scaling}
\end{eqnarray}
These imply the following scaling for the ratio 
\begin{equation}
{r_t\over r_0}={{r_{t*}} \over {r_{0*}}} 
\left({L \over L_*}\right)^{\alpha-1/2}.
\end{equation}
The total mass $M_{\rm tot}(\infty)$ then scales with the luminosity as
\begin{equation}
\,\,M_{\rm tot}(\infty)
\,=\,{2 \pi {\Sigma_0} {r_0} {r_t}}\,=\,{9 \sigma^2 \over
2G}r_t= {9\over 2G}{{\sigma_{*}}^2}{r_{t*}}\left({L \over
L_*}\right)^{\alpha+ 0.5},
\end{equation}
and the mass-to-light ratio $\Upsilon$ is given by
\begin{eqnarray}
{\Upsilon}\, \sim \left( {\sigma_{*}\over 240\,{\rm km~s^{-1}}}\right)^2
   \left( {r_{t*}\over 30\,{\rm kpc}} \right)
   \left( {L\over L_*} \right )^{\alpha-0.5} .
\end{eqnarray}
For $\alpha = 0.5$ the model has constant $\Upsilon$ for all galaxies
(although $\Upsilon$ may be a function of radius inside each galaxy).  If
$\alpha> 0.5$ ($\alpha< 0.5$) then brighter galaxies have a larger
(smaller) $\Upsilon$ than fainter ones.  The parameters that characterize
both the global smooth component and the clumps are optimized using the
observational data (i.e. the positions, magnitudes, geometry of strong
lensing features, and the smoothed shear field) as constraints. We find
that $\alpha = 0.5$, is the favored value for the best-fit mass models and
in fact, $\alpha<0.5$ is ruled out at a very high confidence.

Our goal is to optimally partition the total cluster mass between a smooth
component and the clumps based on the observational data. A
maximum-likelihood method is used to obtain significance bounds on fiducial
parameters that characterize a typical $L_*$ halo in the cluster. The
likelihood function of the estimated probability distribution of the source
ellipticities is maximized for a set of model parameters, given a
functional form of the intrinsic ellipticity distribution measured for the
faint background galaxies.  For each faint galaxy $j$ with measured shape
$\tau_{\rm obs}$, the intrinsic shape $\tau_{S_j}$ can be estimated in the
weak regime by subtracting the lensing distortion induced by the smooth
cluster and galaxy halos,
\begin{eqnarray}
\tau_{S_j} \,=\,\tau_{\rm obs_j}\,-{\Sigma_i^{N_{\rm gal}}}\,
{\gamma_{p_i}}\,-\, \gamma_{c}, 
\end{eqnarray}
where $\Sigma_{i=1}^{N_{c}}\,{\gamma_{p_i}}$ is the sum of the shear
contribution at a given position $j$ from $N_{\rm gal}$ galaxies. This entire
inversion procedure is performed within the {\sc lenstool} utilities developed
by Kneib (1993), which accurately take into account the non-linearities
arising in the strong lensing regime. Using a well-constrained `strong
lensing' model for the inner-regions of the cluster along with the averaged
shear field and assuming a known functional form for $p(\tau_{S})$ from the
field, the likelihood for a guessed model is
\begin{eqnarray}
 {\cal L}({{\sigma_{*}}},{r_{t*}}) = \Pi_{j=1}^{N_{\rm gal}} p(\tau_{S_j}).
\end{eqnarray}
We compute ${\cal L}$ by assigning the median redshift corresponding to the
observed source magnitude for each arclet. The best fitting model
parameters are then obtained by maximizing the log-likelihood function with
respect to the parameters ${\sigma_{*}}$ and ${r_{t*}}$.
Using a bootstrap method, we have verified that while the convergence to
the best-fit model is indeed driven by the brighter cluster galaxies,
no single galaxy dominates in the procedure.

\section{Properties of A\,2218}  

The lens model for the 
mass distribution in the well-studied cluster A\,2218 at $z =
0.17$ is arguably the best constrained case  presently available
(Kneib et al.\ 1996), to which we can 
apply our analysis. We use 
four sets of multiple images, which are all
spectroscopically confirmed 
at $z=0.70$, 1.03, 2.52 and 5.60, in 
conjunction with the background shear field, to obtain the
best-fit parameters for a fiducial $L_*$ galaxy,  $r_t* =
40^{+30}_{-10}$ kpc and $\sigma* = 180^{+15}_{-20}$\,km\,s$^{-1}$ (where the
quoted error bars are at 5$\sigma$ confidence). The mass distribution in
A\,2218 is composed of two clumps centered around the two brightest cluster
galaxies, and the profile inferred from lensing observations and X-ray data
are consistent. Although the mass distribution is bimodal, we note that the
primary clump contributes $\sim 90\%$ of the mass. Nevertheless, we take
the presence of the second clump into account when calculating the
center-of-mass of the cluster. Below, we tabulate the model parameters for
these two components in our best-fit mass model of A\,2218.

\begin{table*}
{\small
\begin{center}
\begin{tabular}{lcccccccc}
\hline\hline\noalign{\smallskip}
${\rm Cluster}$&${z}$&${x}$&{y}&${\epsilon}$&{posn. angle}&${\sigma}$&${r_0}$&${r_t}$\\
&{}&(arcsec)&(arsec)&{}&{(deg)}&(kms$^{-1})$&(kpc)&(kpc)\\
\noalign{\smallskip}
\hline
\noalign{\smallskip}
{\rm Main clump} & ${0.17}$ & {2.0}& {0.3}&${0.3\pm0.05}$&${-13\pm5}$
&{${1070\pm70}$}& ${75\pm10}$ & ${1100\pm 100}$\\
{\rm Second clump} & ${0.17}$ & {$-67.5$}& {3.0}& ${0.2}$ &{$20\pm4$}&
{ ${400\pm30}$}& ${25\pm5}$ & ${600\pm 40}$\\
\noalign{\smallskip}
\hline
\end{tabular}
\end{center}}
\end{table*}

The morphologies and luminosities of the cluster members (which are
crucial for the lensing analysis) are also
well-determined from the high-resolution {\it HST} imaging
of this cluster (Smail et al.\ 2001).
We have used the 40 brightest early type-cluster galaxies in the
maximum-likelihood analysis to obtain the values quoted above.  Their
individual truncation radii were then determined by utilizing the scaling
laws with luminosity in equation~(\ref{eq:scaling}).  Out of the early-type
cluster galaxies selected for the lensing analysis, $\sim 75\%$ are located
inside the core-radius and the rest lie interior to 300\,kpc, within a few
core radii from the cluster center.

There have been several dynamical studies of A\,2218 that yielded
accurate measurements of galaxy velocities relative
to the center of mass of the cluster (Danese, De Zotti \&
di Tullio 1980; Le Borgne et al.\ 1992; Girardi et al.\ 1997; Rakos,
Dominis \& Steindling 2001).  Using the spectro-photometric survey of
Le Borgne et al.\ (1992), we find that 25 of the 40 early-type
galaxies in our analysis have measured redshifts which allow us to
estimate their velocity relative to the center-of-mass of the
cluster. We have then computed the expected truncation radii for these
galaxies in two limiting cases regarding the composition of their dark
matter halos: (i) collisionless dark matter; and (ii) fluid-like dark
matter (see Furlanetto \& Loeb 2002).

In our analysis, we have to additionally assume that the orbits of the
cluster galaxies under consideration are representative of tidally stripped
galaxies. Ghigna et al.\ (1998) have found that most of the strongly
tidally-stripped halos move on nearly radial orbits (since these orbits
penetrate the high density core of the cluster) and that radial orbits are
overabundant by an order of magnitude relative to tangential orbits in the
inner $r\la 500$\,kpc of the cluster. Since we are using the projected
location of cluster galaxies in conjunction with their orbital velocities
at the present time, the nature of the orbits interior to 300\,kpc needs to
be established in order to justify the assumption that these galaxies are
indeed populating representative orbits for efficient tidal
stripping. Using the determined mass profile of the cluster, the
line-of-sight velocity dispersion profile and the number density profile of
the galaxies, and assuming that cluster galaxies are good tracers of the
potential well, we can compute the velocity anisotropy
parameter\footnote{Note that $\beta = 0$ implies isotropic orbits, $0 <
\beta \leq 1$ implies mostly radial orbits and $\beta < 0$ indicates orbits
that are primarily tangential.}, $\beta (r) = (1 -
{{\sigma_{t}^{2}}/{\sigma_{r}^{2}}})$, via the anisotropic Jeans equation
using the approximation of spherical symmetry (for further details see
Natarajan \& Kneib 1996),
\begin{equation}
\frac{d \,(\nu_{g}\,\sigma_{r}^{2})}{dr} + \frac {2 \beta(r) 
\nu_{g} \ \sigma_{r}^{2}}
{r}  =  - \frac {\ G \ M_{\rm tot}(r) \nu_{g}}{r^{2}};
\end{equation}   
where $\nu_{g}(r)$ is three-dimensional galaxy density profile (as derived
from the 2D projected distribution via the Abel transform),
$\sigma_{r}^{2}(r)$ the radial velocity dispersion of the galaxies, and
$M_{\rm tot} (r)$ is the total mass profile.  The 
observed line-of-sight velocity
dispersion profile $\sigma_{los} (R)$ is given by,
\begin{eqnarray}
\frac{1}{2} \, [\,\Sigma_{g}(R) \, \sigma_{los}^{2}(R) ] \, = \,
\int_{R}^{\infty} \frac {r \nu_{g}(r) \sigma_{r}^{2}(r)\, dr}{\sqrt{(r^{2}
- R^{2})}} \,\,[1\, - {\frac {R^{2}}{r^2}} \beta (r)].
\end{eqnarray} 
The above two integro-differential equations can be solved numerically for
$\sigma_{r}^{2}$ and $\beta (r)$. Substituting the best-fit mass model for
A\,2218 we obtain $\beta \approx 0.3$, which indicates that the majority of
orbits are in fact radial and do plunge through the core of the
cluster. Such orbits are likely to facilitate tidal truncation as
assumed in our analysis.

\section{Results}

For collisionless cold dark matter (CDM) the tidal truncation radius,
$r_{t_i}$, for the $i$-th galaxy whose pericenter lies at a radial position
$r_i$ inside the cluster, is defined by the condition that the average
density of the galaxy halo interior to the tidal radius, matches the mean
density of the cluster interior to the galaxy location $r_i$ (Ghigna et
al.\ 1998; Taylor \& Babul 2001):
\begin{equation}
{\langle{\rho_{g_i}}}(r_{t_i})\rangle = \langle \rho_{cl} (r_i)\rangle.
\end{equation}  

For the case of fluid-like dark matter (FDM), the halo truncation process
is modified due to the additional process of ram-pressure stripping. In
this case the truncation radius is set by the condition (Furlanetto \& Loeb
2002),
\begin{equation}
\rho_c (r_i) v_{g_i}^2 = \rho_{g_i}(r_{t})\,\sigma_{g_i}^2,
\end{equation} 
where $v_{g_i}$ is the velocity component of the $i$-th galaxy
relative to the cluster center-of-mass, $\rho_{g_i}(r_{t})$ is the
density of that galaxy halo at its 3D truncation radius $r_t$, and
$\sigma_{g_i}$ is the internal velocity dispersion of that
galaxy\footnote {Note that while the truncation radii inferred from
the lensing observations are projected 2D values, those calculated
from equations (7) and (8) are in 3D. However, because the PIEMD mass
density profile is $\rho\propto (r^{2}+r_0^2)^{-1}(r^{2}+r_t^2)^{-1}$,
both truncation radii are equivalent.}. In our analysis, we have
conservatively used only the measured component of the velocity of
each galaxy (relative to the cluster center) along
line-of-sight. Since the full (3D) galaxy velocity can only be larger
than our adopted value, our constraint on the truncation radii of
galaxies in the FDM case should be regarded as conservative upper
limits.  The considerations quoted above ignore any further truncation
that is likely to be caused by thermal or turbulent heating (Gnedin \&
Ostriker 2001).  Using the above expressions, we derived the
truncation radii for the 25 early-type galaxies in A\,2218 in the two
limiting cases.  The results are plotted in Figure~2.  The observed
distribution of truncation radii from the lensing data analysis (solid
triangles) is compared to the radii expected in the collisionless
limit (solid squares) and the fluid limit (solid circles).  We find
that FDM models are ruled-out at a confidence greater than 5$\sigma$,
while collisionless dark matter models are in excellent agreement with
the observationally determined values.  We find no correlation between
the truncation radii and the orbital velocities for the cluster
members in A\,2218, as would be expected if ram pressure stripping had
been a significant dynamical process in operation.

The association of the retrieved model radius $r_t*$ from the
lensing analysis with the tidally truncated radius is strengthened
through the following observed correlation. For a galaxy described by
a PIEMD mass profile with velocity dispersion $\sigma_{0_i}$ and whose
orbital pericenter crosses the cluster core where the density is
$\rho_0$, the tidal truncation radius scales as
\begin{equation}
{r_{t}}\,\propto\,{\sigma_{0_i}}\,{\rho_0}^{-\frac{1}{2}}.
\end{equation}
Deriving the best-fit values of $\rho_0$ from the strong lensing
models and $\sigma_{0_i}$ from the maximum-likelihood analysis, we
find that the above simple scaling adequately reproduces the observed
trends both for the truncation radius and the mass of a
fiducial halo in the five {\it HST} clusters studied 
in NKS02 (ranging in redshift from 
$z = 0.17$--0.58).  The trends
seen in halo size $r_t^*$ with redshift from the lensing study of
NKS02 are also in good agreement with the theoretical expectation from
numerical simulations (Ghigna et al.\ 1998).  High resolution
collisionless N-body simulations of cluster formation and evolution,
predict that the dominant interaction is between the global cluster
tidal field and individual galaxies in the redshift interval $z\sim
0$--0.5 (Ghigna et al.\ 1998; Moore et al.\ 1996).  

\section{Conclusions and Discussion}

We have used gravitational lensing data to constrain the truncation radii
of galactic dark matter halos in the cluster A\,2218.  Figure~2 shows that
the inferred truncation radii are consistent with the tidal radii expected
for collisionless dark matter, but rule-out fluid-like dark matter for
which ram pressure stripping is effective.

The transition between the collisionless and collisional regimes is set
by the ratio between the mean-free-path of dark matter particles,
$\lambda$, and the radius, $r_t$, of the galaxy halos under
consideration. This ratio is given by,
\begin{equation}
{\lambda\over r_t} \approx {m_p \over \sigma_p \Sigma(r_t)} ,
\end{equation}
where $\sigma_p/m_p$ is the collisional cross-section per unit particle
mass and $\Sigma (r_t)$ is the surface mass density of a galaxy halo at its
truncation radius. The fluid regime is obtained for ${\lambda\over r_t}\la
1$.  The characteristic surface mass densities of the galaxies in A\,2218 can
be directly inferred from the analysis of the lensing data.  We find that
for an $L_\star$ galaxy, $\Sigma_\star(r_t)\approx 0.024 ~{\rm
g~cm^{-2}}$. Since the fluid regime is ruled out, we exclude all values of
$\sigma_p/m_p \ga 42~{\rm cm^2~g^{-1}}$.

Our constraints are complementary to those derived by Gnedin \&
Ostriker (2001) from considerations of thermal conduction in the
mildly collisional regime.  These authors exclude the regime $0.3\la
\sigma_p/m_p\la 10^4$\,cm$^2$\,g$^{-1}$, based on the consideration
that cluster ellipticals will otherwise deviate from the fundamental
plane beyond the observed scatter. Our new constraint allows us to
rule out the high cross-section regime and hence we conclude that
$\sigma_p/m_p\la 0.3$\,cm$^2$\,g$^{-1}$ as an absolute upper limit.
Dark matter cross-sections higher than this upper limit were
postulated by Spergel \& Steinhardt (2000) in order to reconcile
problems that cold dark matter models possess when compared to
observational data (such as the abundance of galactic sub-halos and
the slope of the inner mass density profile of galaxies; see further
discussion in Dave et al.\ 2001; Yoshida et al.\ 2000; Stoehr et al.\
2002; Miralda-Escude 2002).

\acknowledgments

PN acknowledges support from a Trinity College Research Fellowship.
This work was supported in part by NSF grants AST-9900877, AST-0071019
for AL. JPK thanks the CNRS and the TMR-Lensing collaboration and IRS
the Royal Society and the Leverhulme Trust for support.

\newpage
 
\begin{figure}
\centerline{\psfig{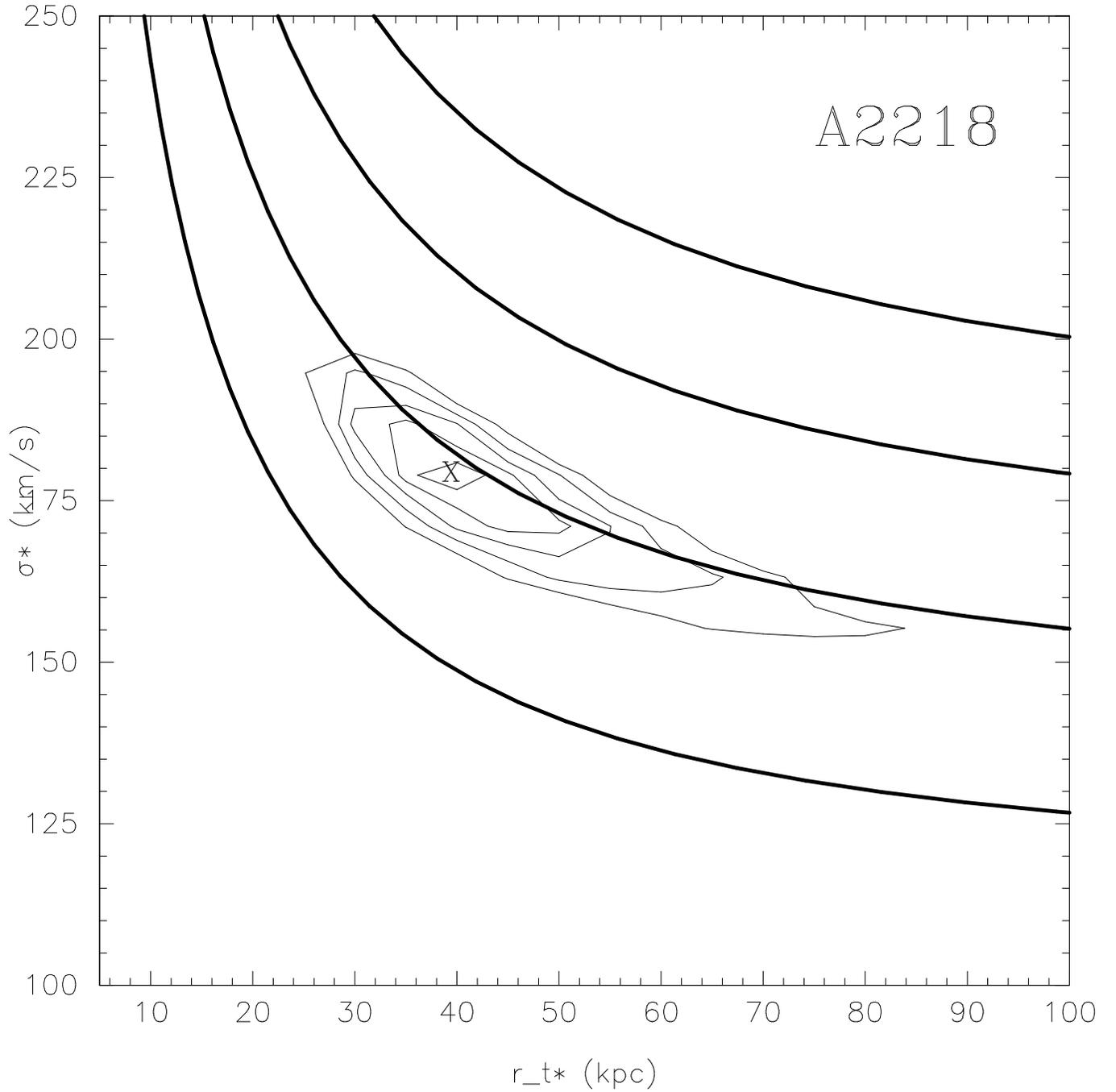}}
\caption{Results of maximum-likelihood analysis of lensing data for the
cluster A\,2218 (NKS02).  The contour plot shows the best-fit values 
for the model
parameters $\sigma_*$ and $r_{t*}$, which are the central velocity
dispersion and truncation radius for a typical $L_*$ galaxy in the cluster.
The likelihood contours are plotted in intervals of $1-\sigma$ starting
from the inside out. The thick open curves are lines of constant
enclosed mass.
}
\end{figure}

\begin{figure}
\plotone{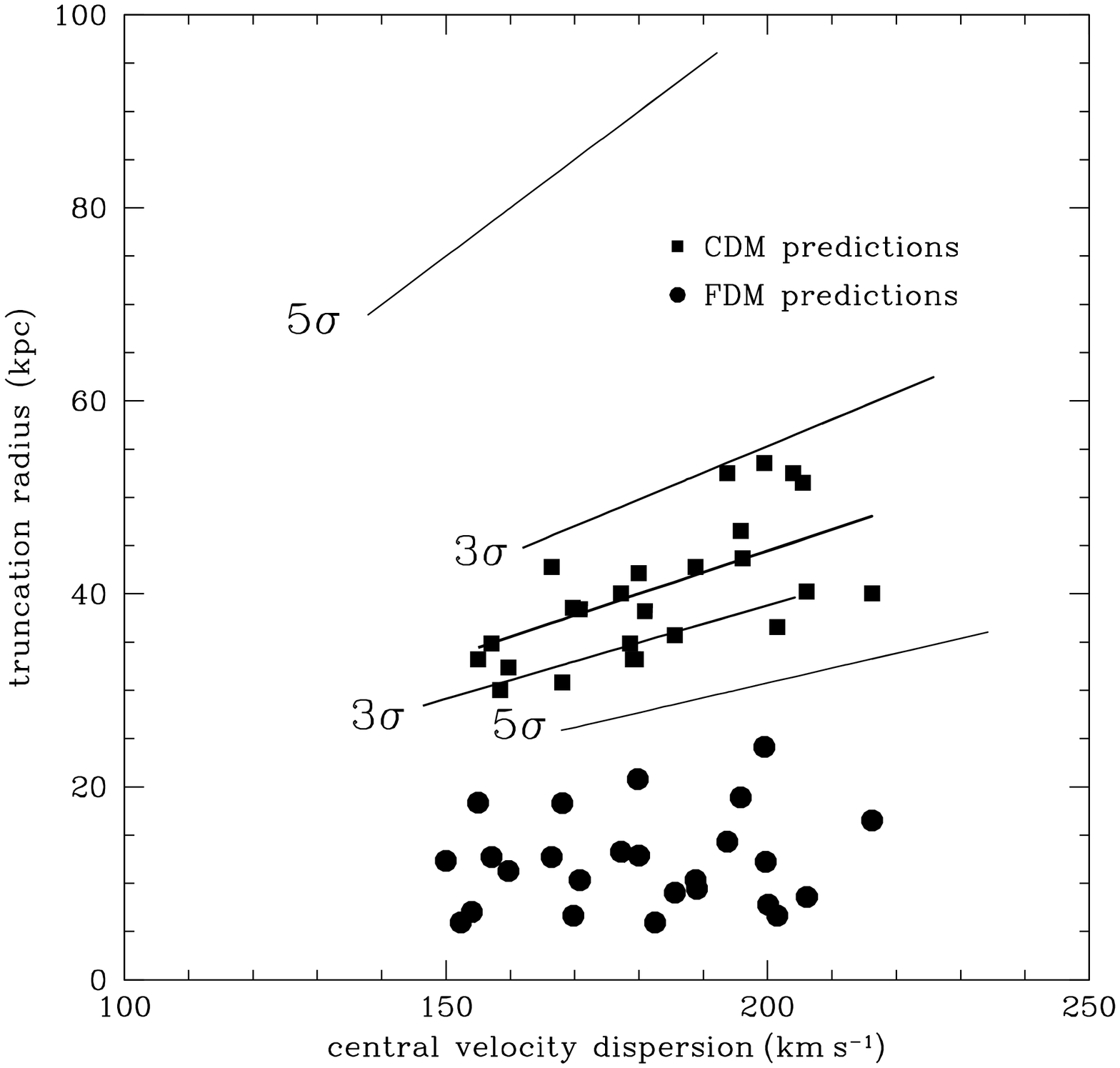}
\caption{Distribution of truncation radii as inferred from the lensing
analysis of 25 galaxies superimposed on the scaling relation (thick solid
line) in the cluster A\,2218. The 3-$\sigma$ and 5-$\sigma$ lines denote the
corresponding confidence levels in the parameters $r_{t*}$, and $\sigma^*$
obtained for a typical $L_*$ galaxy in this cluster (derived from the
confidence level contours in Figure~1). The expected distribution of tidal
radii for fluid-like dark matter (FDM; solid circles) and collisionless
cold dark matter (CDM; solid squares) are also shown. Note that the linear
relation between truncation radius and velocity dispersion for the values
inferred from lensing is a consequence of the assumed scaling laws with
galaxy luminosity [equation (2)].}
\end{figure}


\begin{references}

Binney, J., \& Tremaine, S., 1987, Galactic Dynamics, (Princeton: Princeton
U. Press), Ch. 7

Brainerd, T., Blandford, R., \& Smail, I., 1996, ApJ, 466, 623
 
Couch, W.J., Barger, A.J., Smail, I., Ellis, R.S., Sharples,
R.M., 1998, ApJ, 497, 188

Danese, L., de Zotti, G., \& di Tullio, G. 1980, ApJ, 82, 322

Dave, R., Spergel, D., Steinhardt, P. J., \& Wandelt, B. 2001,
ApJ, 547, 574

Dell' Antonio, I., \& Tyson, J. A. 1996, ApJ, 473, L17.

Fischer, P., et al. 2000, AJ, 120, 1198 
 
Furlanetto, S., \& Loeb, A. 2002, ApJ, 565, 854

Ghigna, S., Moore, B., Governato, F., Lake, G., Quinn, T., \& Stadel, J.
1998, MNRAS, 300, 146

Girardi, M., et al. 1997, ApJ, 490, 56

Gnedin, O., \& Ostriker, J. P. 2001, ApJ, 561, 61

Griffiths, R. E., Casertano, S., Om, M., \& Ratnatunga, K. U.
1996, MNRAS, 282, 1159 

Kassiola, A., \& Kovner, I. 1993, ApJ, 417, 474

Kneib, J-P., Ellis, R. S., Couch, W., Smail, I. R., \& 
Sharples, R. 1996, 471, 643

Le Borgne, J-F., Pello, R., \& Sanahuja, B. 1992, A\&AS, 95,  
87

McKay, T. et al. 2002, ApJ, submitted (astro-ph/0108013) 

Miralda-Escude, J. 2002, ApJ, 564, 1019

Moore, B., Katz, N.,  Lake, G., Dressler, A., \& 
Oemler, A. 1996, Nature, 379, 613

Natarajan, P., \& Kneib, J-P. 1996, MNRAS, 283, 1031

Natarajan, P., \& Kneib, J-P. 1997, MNRAS, 287, 833 

Natarajan, P., Kneib, J-P., Smail, I., \& Ellis, R. S. 1998, ApJ, 499, 600

Natarajan, P., Kneib, J-P., \& Smail, I 2002, ApJ, submitted 

Rakos, K., Dominis, D., \& Steindling, S. 2000, A\&A, 369,750 

Smail, I., et al. 1997, ApJS, 110, 213  

Smail, I., et al. 2001, MNRAS, 323, 839 

Spergel, D., \& Steinhardt, P. J. 2000, Phys. Rev. Lett., 84, 17, 
3760

Stoehr, F., et al. 2002, preprint, astro-ph/0203342 

Taylor, J. E., \& Babul, A. 2001, ApJ, 559, 716

Yoshida, N., Springel, V., White, S. D. M., \& Tormen, G. 2000, ApJ,
535, L103 

Wilson, G., Kaiser, N., Luppino, G. A., \& Cowie, L. L. 2001, ApJ, 555, 572
\end{references}
\end{document}